**Spin-Polarized Ground States and Ferromagnetic Order Induced by Low-Coordinated Surface Atoms and Defects in Nanoscale Magnesium Oxide**

Takashi Uchino*

*Department of Chemistry, Graduate School of Science, Kobe University, Nada, Kobe 657-8501, Japan*

Toshinobu Yoko

*Institute for Chemical Research, Kyoto University, Uji, Kyoto 611-0011, Japan*

We investigate the effect of low-coordinated surface atoms on the defect-induced magnetism in MgO nanocrystallites using hybrid density functional theory calculations. It has been demonstrated that when Mg vacancies are introduced at the corners of cube-like MgO clusters, a magnetic state ($S \geq 1$) becomes lower in total energy than the nonmagnetic singlet state by 1–2 eV, resulting in the spin-polarized ground state. The spin density is not only located at the surrounding O atoms neighbor to the corner Mg vacancy ($V_{Mg}^{C}$) site but is also extended to the distant (~1 nm or longer) low-coordinated surface O atoms along the <110> directions. This directional spin delocalization allows a remote $V_{Mg}^{C} - V_{Mg}^{C}$ interaction, eventually leading to a spontaneous long-range ferromagnetic interaction.

*Author to whom correspondence should be addressed: uchino@kobe-u.ac.jp

Recent progress in materials synthesis has led to a variety of nanostructures whose optical, electrical and magnetic properties are very different from those of bulk structures. This also reveals anomalous aspects of well understood phenomena of condensed matter physics. One of the interesting examples of the anomalies found in nanoscale oxides is the so-called "$d^0$ ferromagnetism [1]," which is referred to as non-negligible ferromagnetic moments observed in closed shell oxide systems, e.g., $HfO_2$ [2], $TiO_2$ [3,4], $CeO_2$ [5], ZnO [5], $Al_2O_3$ [5], and MgO [5–8], containing virtually no atoms with partially filled $d$ or $f$ shells. This intriguing ferromagnetism occurs even at temperatures well above room temperature and seems to exist outside the conventional $m$-$J$ paradigm [1,9,10], where $m$ represents the magnetic moment and $J$ represents the exchange coupling of electron spins. Thus, $d^0$ ferromagnetism is observed almost universally in nanoscale oxides and is believed to result from a certain defective state [7–13]. However, a simple defect-based model cannot account for the observed ferromagnetic behavior because of a rather localized nature of the relevant defect orbitals, which will be insufficient to induce long range ferromagnetic order, as has been shown by recent density functional theory (DFT) calculations including electronic correlation effects beyond standard (semilocal) DFT functionals [14–16].

When we say "defects" in oxides, we usually mean "atomic defects," e.g., vacant atom sites, interstitial atoms, or substitution of a foreign atom for a normal one, in the bulk. Thus, previous theoretical calculations have been performed mainly on the atomic defects in the bulk using supercell methods [7,14,15,17]. However, the atoms located at the surface of crystals can also be regarded as structural defects or imperfections because of the reduced number of coordination bonds. As for nanoscale oxides, the possible effect of these surface atoms may not be completely neglected. Furthermore, it has been demonstrated that the low-coordinated surface sites such as terraces, edges, and corners are more stable locations for atomic defects than the bulk sites because the formation energy of atomic defects will decrease with decreasing the number of surrounding atoms [18,19]. This allows one to expect that these low-coordinated surface states could potentially contribute to the generation of defect-related ferromagnetism in nanoscale oxide, as suggested by recent observations [6,8,20]. However, the possible surface effects on the defect related magnetism have not been well investigated theoretically [17].

We hence carry out a series of quantum chemical DFT calculations on isolated MgO clusters with surface atomic defects such as O and Mg vacancies. Such isolated cluster models are not appropriate to investigate the electronic structure of the bulk system and the extended surface of crystals. When materials are reduced to the nanoscale, however, their structure and properties can deviate from the bulk or extended surface case. Accordingly, a real space, rather than reciprocal space description, will become more effective, implying that isolated cluster models are simple but reasonable models especially for nanocrystals.

As for MgO clusters consisting of more than ~50 atoms, it has been demonstrated that the cubic rock-salt structure thermodynamically dominates the energy landscape [21]. We hence employed several top-down clusters based on cuts from the cubic rock-salt structure as representative models of nanometer-sized MgO crystals. We first consider a (4×4×4)-atom block of stoichiometric $Mg_{32}O_{32}$ cluster consisting in total of 64 atoms, as shown in Fig. 1(a). To evaluate the effect of a surface Mg (or O) vacancy on the stable spin state we intentionally removed one Mg (or O) atom at one of the corner sites of the 4×4×4 cluster. We then performed full geometry optimization for these Mg-deficient ($Mg_{31}O_{32}$) and O-deficient ($Mg_{32}O_{31}$) clusters, starting from ideal cubic configurations, at the spin-restricted singlet ($S=0$) and spin-unrestricted triplet ($S=1$) spin states without imposing any structural constraints. All the DFT calculations in this work were carried out using the gradient corrected Becke's three parameters hybrid exchange functional [22] in combination with the correlation functional of Lee, Yang, and Parr [23] (B3LYP) with the GAUSSIAN-09 code [24]. It has previously been shown that such a hybrid DFT functional is useful to correct the self-interaction problem [25], which often leads to misleading conclusions with regards to hole localization and the resulting magnetic characteristics of the system [16,26]. Mulliken's population analysis was conducted to calculate the spin densities of the clusters at B3LYP/6-31G(d) level. The stability of the resulting optimized clusters was evaluated in terms of the total energy along with the atomization (AE), which is defined as the energy necessary to dissociate the $Mg_mO_n$ cluster into neutral atoms ($m$Mg + $n$O), namely, AE=$mE$(Mg) + $nE$(O) − $E(Mg_mO_n)$, where $E$(X) represents the total energy of the system $X$. The AE is useful to evaluate the stability of the clusters with the same dimension but different compositions and spin states.

We found that irrespective of the assumed spin state, the starting cubic configuration is almost retained for all the clusters employed after full geometry optimization although slight outward atomic displacements with respect to the respective vacancy sites were seen. It should be noted, however, that the stability of the spin state varies depending on the type of defect included in the cluster (see Table I). The lower energy spin state of the O-deficient ($Mg_{32}O_{31}$) cluster is the singlet ($S=0$) state, in agreement with the results of supercell calculations [17,27]. As for the Mg-deficient ($Mg_{31}O_{32}$) cluster, however, the triplet ($S=1$) state is substantially lower in total energy than the singlet ($S=0$) state by ~1.3 eV. Such a large triplet-singlet energy gap has not been obtained for the Mg vacancy in the supercell-based model, where the triplet state is generally almost degenerate with the singlet state [17]. Note also that the Mg-deficient cluster in the triplet state yields the largest AE value and hence can be regarded as the most stable defect configuration among the clusters shown in Table I.

The spin-polarized ground state of the Mg-deficient cluster can be interpreted in terms of the molecular orbital diagrams shown in Fig. 1. As for the Mg-deficient cluster in the singlet state, the lowest unoccupied molecular orbital (LUMO), which is characterized basically by $2p$ orbitals of O atoms, is only slightly higher in energy than the highest occupied molecular orbital (HOMO) by ~0.5 eV [see Fig. 1(c)]. This contrasts with the case of the O-deficient cluster in the singlet state, where the LUMO is higher in energy than the HOMO by ~1.7 eV [see Fig. 1(b)]. It hence follows that spontaneous spin polarization is expected to occur in the Mg-deficient cluster because of a small HOMO-LUMO gap, which costs less energy to flip a spin, leading to the spin polarized ground state [see Fig. 1(d)].

We next investigate the spin-magnetization density, which is defined as the local density difference between the spin-up and spin-down states, of the Mg-deficient cluster in the triplet state (see Fig. 2). One sees from Fig. 2 that most of the spin is distributed over the oxygen atoms adjacent to the corner Mg vacancy. This feature is basically in agreement with that calculated previously for the Mg vacancy introduced into an MgO supercell [17]. In the present Mg-deficient cluster, however, a non-negligible spin density is further spread out of the nearest neighbor oxygen atoms along the <110> directions on the {100} surfaces. We also found that the value of the Mulliken atomic spin density of the surface Mg and subsurface O and Mg atoms are below 0.0005, indicating that the defect derived spin density is preferentially distributed over the low-coordinated surface O atoms.

It is hence interesting to investigate whether the defect-induced spin polarization can couple ferromagnetically to each other. This coupling may induce a higher spin ground state or a ferromagnetic ordering. To investigate the effect of the inter-defect distance between a pair of Mg vacancies on the spin state we employed a series of rocksalt-type clusters consisting of (6×4×3)-, (7×4×3)-, and (8×4×3)-atom blocks and removed two corner Mg atoms located in the same (100) plane of the respective clusters, resulting in the $Mg_{34}O_{36}$, $Mg_{40}O_{42}$, $Mg_{46}O_{48}$ clusters [see the inset of Figs. 3(a)–3(c)]. Full geometry optimizations have been performed as well for this series of Mg-deficient clusters at the B3LYP/6-31G(d) level by assuming various spin states, i.e., $S$=0, 1 and 2. The inter-vacancy distances for the optimized $Mg_{34}O_{36}$, $Mg_{40}O_{42}$, $Mg_{46}O_{48}$ clusters are estimated to be ~12.0, ~12.5, and ~15.5 Å, respectively, and the relative total energies of the spin-polarized ($S$=1 and 2) sates with respect to the spin-unpolarized closed-shell ($S$=0) state are summarized in Table II.

One sees from Table II that all these Mg-deficient clusters are characterized by the same the energetic ordering of different spin states, namely, $E_q<E_t<E_s$, where $E_q$, $E_t$, and $E_s$ represent the total energy of the spin quintet ($S$=2), triplet ($S$=1) and singlet ($S$=0) states, respectively. The spin $S$=2 and 0 states are separated by an energy of ~2.6 eV, implying the existence of a robust ferromagnetic coupling between a pair of distant (~ 1 nm) Mg vacancies. Table II also shows that the spin triplet state is only slightly higher in energy (several milli-electron volts) than the spin quintet state, suggesting the degeneracy of these two spin polarized states or the coexistence of ferromagnetic and antiferromagnetic coupling. We should note, however, that an unrestricted DFT (UDFT) single determinant is not an eigenfunction of the total spin operator $S^2$ and inherently has the spin impurity problem [28]. One sees from Table II that the spin-squared expectation values $<S^2>$ for the spin quintet ($S$=2) state are around 6.02−6.03, which is close to the ideal value $S(S+1)$=6, even before annihilation of the first spin contaminant. In the case of the triplet state, however, the $<S^2>$ values are ~3.02 and ~2.05 before and after annihilation of the first spin contaminant, respectively [29]. This indicates that the calculation for the triplet state is affected by mixing with a higher spin (possibly quintet) state.

Figures 3(a)-3(c) show the spin-magnetization density of the clusters with two corner Mg vacancies in the spin quintet state. In the 6×4×3-based $Mg_{34}O_{36}$ cluster, the magnetic interaction between the two Mg vacancies can be explicitly recognized, showing a delocalized and extended nature of the spin polarization of surface O atoms aligned along the <110> directions. The degree of inter-vacancy interaction appears to decrease with increasing the distance between a pair of Mg vacancies, but the directional spin delocalization over the surface O atoms still survives even in the 8×4×3-based $Mg_{46}O_{48}$ cluster. It is hence most likely that the surface O atoms indirectly but inherently contribute to the formation of long-range magnetic order or the ferromagnetic percolation in MgO nanocrystals.

In summary, we have shown from a series of DFT calculations on MgO nanoclusters that a Mg corner vacancy can induce a delocalized spin distribution over several neighboring surface O atoms along the <110> directions [30]. This directional spin delocalization enables a pair of distant (~1 nm or longer) Mg vacancies to interact ferromagnetically, resulting in the spin polarized ground state. These results allow us to suggest that the low-coordinated surface atoms are prerequisite for long-range ferromagnetic interaction, hence providing a delocalized mediating state or a percolation network for defect-related ferromagnetism in nanoscale oxides.

We thank the Supercomputer System, Institute for Chemical Research, Kyoto

University, for providing the computer time to use the SGI UV 100 supercomputer.**References and Notes**

[1] J. M. D. Coey, Solid State Sci. **7**, 660 (2005).

[2] M. Venkatesan, C. B. Fitzgerald, and J. M. D. Coey, Nature (London) **430**, 630 (2004).

[3] S. D. Yoon, Y. Chen, A. Yang, T. L Goodrich, X. Zuo, D. A Arena, K. Ziemer, C. Vittoria and V. G Harris, J. Phys.:Condens. Matter **18**, L355 (2006).

[4] N. H. Hong, J. Sakai, N. Poirot, and V. Brizé, Phys. Rev. B **73**, 132404 (2006).

[5] A. Sundaresan, R. Bhargavi, N. Rangarajan, U. Siddesh, and C. N. R. Rao, Phys. Rev. B **74**, 161306(R) (2006).

[6] J. Hu, Z. Zhang, M. Zhao, H. Qin, and M. Jiang, Appl. Phys. Lett. **93**, 192503 (2008).

[7] C. Martínez-Boubeta *et al.*, Phys. Rev. B **82**, 024405 (2010).

[8] B. M. Maoz, E. Tirosh, M. Bar Sadan, and G. Markovich, Phys. Rev. B **83**, 161201(R) (2011).

[9] M. Stoneham, J. Phys.: Condes. Matter **22**, 074211 (2010).

[10] S. B. Ogale, Adv. Mater. **22**, 3125 (2010).

[11] J. M. D. Coey and S. A. Chambers, MRS Bull. **33**, 1053 (2008).9

singlet state using a broken-symmetry spin unrestricted formalism where opposite spin electrons have separate Kohn-Sham orbitals that are allowed to polarize. We found that the UDFT singlet state results in even larger and anomalous values of $S^2$ for all the clusters with two Mg corner vacancies; the $S^2$ values were ~2.0 and ~4.3 before and after spin annihilation. Thus, the open-shell singlet state cannot be regarded as a realistic spin state.

30. In our recent paper [T. Uchino and T. Yoko, Phys. Rev. B **85**, 012407 (2012)], we have carried out a series of DFT cluster calculations for non-defective cubic (but Mg-deficient) $Mg_{13}O_{14}$ (3×3×3) and $Mg_{62}O_{63}$ (5×5×5) clusters. Although these 3×3×3- and 5×5×5-clusters do not include an Mg vacancy, we found that the spin triplet state is lower in total energy than the spin singlet state. Furthermore, the spin density was extended over the surface oxygen atoms along the <110> directions although this feature was not explicitly mentioned by the present authors at the time of original publication. It is hence most likely that the Mg-deficient MgO nanocrystals, with or without Mg vacancies, commonly exhibit a ferromagnetic interaction mediated by surface oxygen atoms located along the <110> directions.

**Table I.** Singlet-triplet energy differences $\Delta E$ and atomization energies AE for the fully optimized O-deficient ($Mg_{32}O_{31}$) and Mg-deficient ($Mg_{31}O_{32}$) clusters at the B3LYP/6-31G(d) level.

| cluster | spin state | $\Delta E^a$ (eV) | AE (eV) |
|---|---|---|---|
| $Mg_{32}O_{31}$ | singlet | – | 270.588 |
|  | triplet | 1.01 | 269.583 |
| $Mg_{31}O_{32}$ | singlet | – | 269.409 |
|  | triplet | −1.30 | 270.712 |

$^a$ $\Delta E = E_{triplet} - E_{singlet}$. $E_{triplet}$ and $E_{singlet}$ represent the total energy of the spin-triplet and spin-singlet states, respectively.

Table II. Electronic properties of the Mg-deficient clusters with two corner Mg vacancies optimized at the B3LYP/6-31G(d) level in the respective spin states.

| cluster | spin state | $\Delta E^a$ (eV) | AE (eV) | $<S^2>^b$ |
|---|---|---|---|---|
| $Mg_{34}O_{36}$ | singlet | – | 295.868 | 0 |
| (6×4×3) | triplet | −2.679 | 298.547 | 3.020 (2.045) |
|  | quintet | −2.682 | 298.550 | 6.023 (6.000) |
| $Mg_{40}O_{42}$ | singlet | – | 350.281 | 0 |
| (7×4×3) | triplet | −2.670 | 352.951 | 3.024 (2.045) |
|  | quintet | −2.673 | 352.954 | 6.024 (6.000) |
| $Mg_{46}O_{48}$ | singlet | – | 404.873 | 0 |
| (8×4×3) | triplet | −2.687 | 407.560 | 3.024 (2.049) |
|  | quintet | −2.690 | 407.563 | 6.024 (6.000) |

[a]Triplet or quintet state energy minus closed-shell singlet state energy. [b]The values in parentheses are $S^2$ values after annihilation of the first spin contaminant.

**Figure captions.**

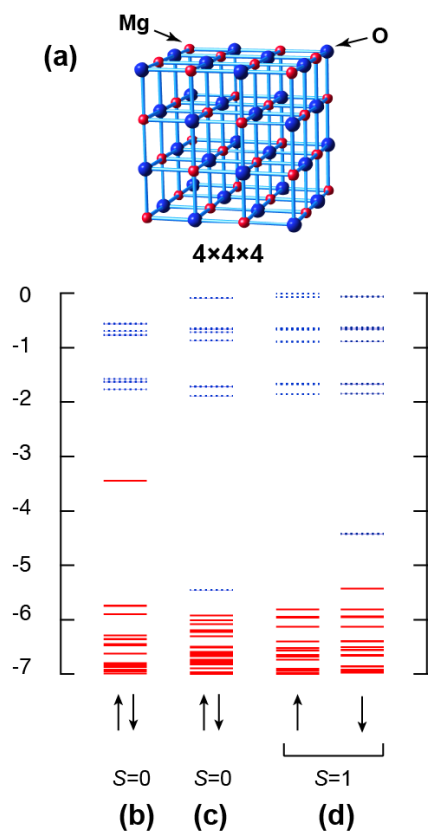

**Figure 1.** (Color online) (a) A (4×4×4)-atom block of the $Mg_{32}O_{32}$ cluster. A corner Mg (or O) atom, indicated by an arrow, is removed to create an Mg-(or O-) deficient cluster. Molecular orbital energy-level diagrams of occupied molecular orbitals (solid lines) and unoccupied molecular orbitals (dotted lines) calculated for (b) the O-deficient $Mg_{32}O_{31}$ cluster in the spin-singlet state, (c) the Mg-deficient $Mg_{31}O_{32}$ cluster in the spin-singlet state, (d) the Mg-deficient $Mg_{31}O_{32}$ cluster in the spin-triplet state.

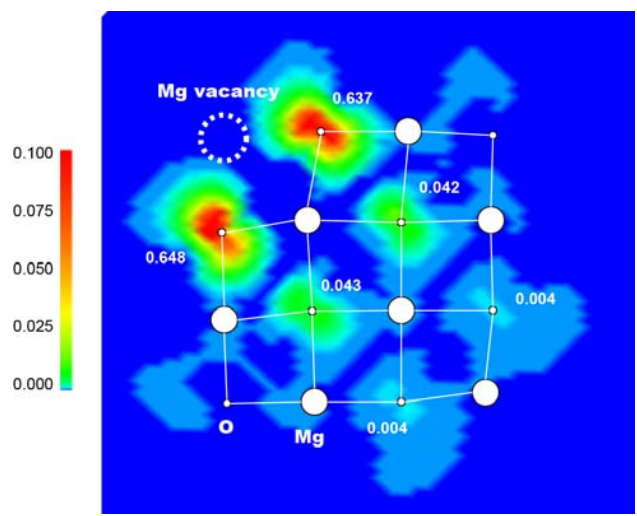

**Figure 2**. (Color online) Spin density map on one of the (100) planes for the $Mg_{31}O_{32}$ cluster in the spin-triplet state. Large and small circles represent Mg and O atoms, respectively. The values indicated are Mulliken atomic spin densities; the values below 0.001 are omitted.

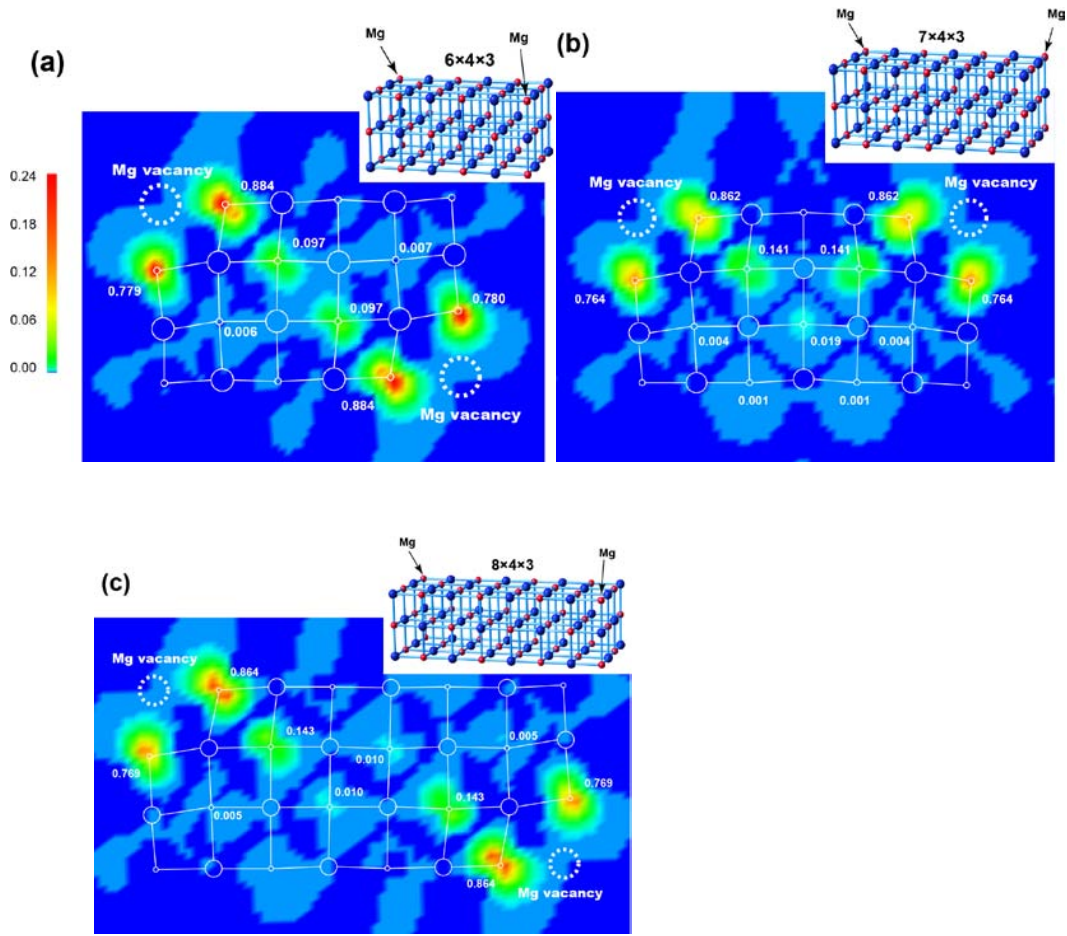

**Figure 3**. (Color online) Spin density map on the (100) plane containing two corner Mg vacancies calculated for the (a) $Mg_{34}O_{36}$ (b) $Mg_{40}O_{42}$ and (c) $Mg_{46}O_{48}$ clusters in the spin-quintet state. Large and small circles represent Mg and O atoms, respectively. The values indicated are Mulliken atomic spin densities; the values below 0.001 are omitted. Insets show the (6×4×3)-, (7×4×3)-, and (8×4×3)-atom blocks, in which the positions of the two corner Mg sites removed are indicated by arrows.